# Preserving the Stoichiometry of Triple-Cation Perovskites by Carrier-Gas-Free Antisolvent Spraying

*Oscar Telschow[1,2], Miguel Albaladejo-Siguan[1,2], Lena Merten,[3] Alexander D. Taylor[1,2], Katelyn P. Goetz[1,2], Tim Schramm[1,2], O. V. Konovalov[4], M. Jankowski[4], Alexander Hinderhofer,[3] Fabian Paulus,[2] Frank Schreiber[3] and Yana Vaynzof[1,2]\**

1. Integrated Center for Applied Physics and Photonic Materials, Technische Universität Dresden, Nöthnitzer Straße 61, 01187 Dresden, Germany

2. Center for Advancing Electronics Dresden (cfaed), Technische Universität Dresden, Helmholtzstraße 18, 01069 Dresden, Germany

3. Institut für Angewandte Physik, Universität Tübingen, 72076 Tübingen, Germany.

4. ESRF, The European Synchrotron, 71 Avenue des Martyrs, CS 40220, 38043 Grenoble Cedex 9, France

**Corresponding Author**

\* E-mail: yana.vaynzof@tu-dresden.de

ABSTRACT

The use of antisolvents during the fabrication of solution-processed lead halide perovskite layers is increasingly common. Usually, the antisolvent is applied by pipetting during the spin-coating process, which often irreversibly alters the composition of the perovskite layer, resulting in the formation of $PbI_2$ at the surface and bulk of the perovskite layer. Here, we demonstrate that by applying the antisolvent via carrier-gas free spraying, the stoichiometry of the perovskite layer is far better preserved. Consequently, the photovoltaic performance of triple cation photovoltaic devices fabricated in an inverted architecture is enhanced, mainly due to an increase in the open-circuit voltage. By exploring different volumes of antisolvent, we show that spraying as little as 60 µL results in devices with power conversion efficiencies as high as 21%. Moreover, solar cells with sprayed antisolvent are more stable than those fabricated by pipetting the antisolvent.

**TOC GRAPHICS**

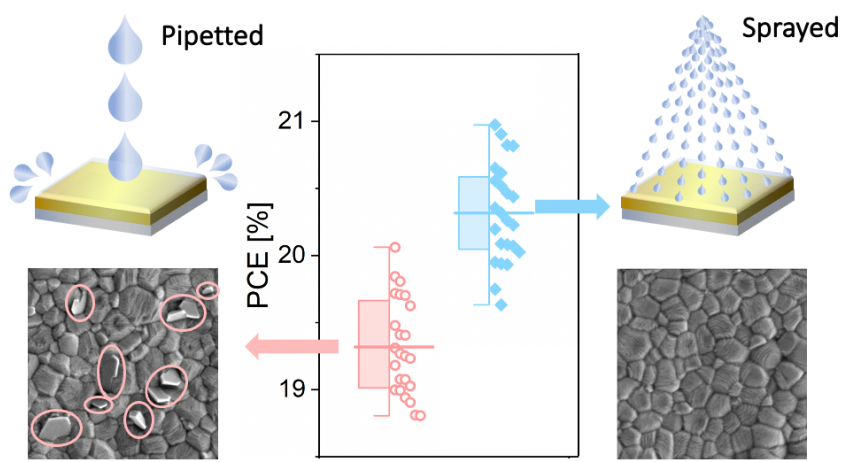

The remarkable progress in the development of photovoltaic devices based on lead halide perovskites makes them particularly promising for industrial integration.[1-3] However, the most efficient devices are still fabricated using the so-called 'solvent-engineering' method,[4-5] in which an antisolvent is used to trigger the crystallization of the perovskite layer during the spin-coating of the perovskite precursor solution (**Figure 1a**). This has led to extensive optimization of the antisolvent application step, including, among others, parameters such as choice of solvent,[6-7] its volume,[8-9] temperature[10-11] and application timing,[12] which has led to the demonstration of high-efficiency photovoltaic devices. Despite this optimization, there are several drawbacks of the current method of antisolvent application. First of all, depositing the antisolvent by pipetting it onto the wet perovskite films commonly results in the use of relatively large volumes of antisolvent (100s of μL) to ensure the coating of the surface of the entire sample.[13-14] Consequently, the large volume of available antisolvent often leads to interactions with the underlying wet perovskite layer that can irreversibly modify its stoichiometry. For example, even a minute solubility of the perovskite precursors (e.g. organic halides) in the antisolvent may lead to their diffusion across the liquid-liquid interface (between the wet perovskite film and the antisolvent) and their removal from the perovskite layer.[15-16] This might explain why perovskite films made from a perfectly stoichiometric solution result in a non-stoichiometric composition, with excess $PbI_2$ observed in either scanning-electron microscopy (SEM) or X-ray diffraction (XRD) measurements.[15,17-18] Furthermore, the use of large volumes of antisolvent raises concerns about the scalability of deposition via the solvent engineering method,[19-20] as well as the environmental impact of these solvents.[21-22] These drawbacks motivate the search for alternative methods for the application of antisolvents.

Spraying was investigated as an alternative method for dispensing the antisolvent by several researchers aiming in particular to improve the homogeneity of perovskite samples. Early work by Ye *et al*. showed that spraying the antisolvent using dry air as a carrier gas led to the formation of smoother, more uniform films, while dispensing the antisolvent by pipetting resulted in coffee-ring effects.[23] While the authors did not specify the volume of the sprayed/pipetted antisolvent, they reported that the optimal devices were obtained by continuously spraying the wet perovskite film for 19-24 s using a high carrier-gas flow of 90 L/h. These parameters suggest that the volume of sprayed antisolvent was relatively high. Similarly, improved surface coverage and homogeneity were observed when spraying large area perovskite films using a carrier-gas (unspecified) assisted spraying at 60 psi for 3 s.[24] A more recent study by Lee *et al*. has explored spraying mixed antisolvents (acetonitrile/chlorobenzene at various ratios) as a method to suppress the inhomogeneities the authors observed when dispensing the mixed antisolvents by pipetting.[25] Despite these promising results, carrier-gas-assisted spraying has several limitations. The use of a carrier-gas might introduce additional artifacts, as the flow of gas onto the wet perovskite film might impact the drying dynamics. Furthermore, past results have shown that carrier-gas assisted spraying requires longer application times (varying from 3 to 24 s of continuous spraying), which adds to the antisolvent consumption and might irreversibly affect the composition of the perovskite layer.

In this work, we explore carrier-gas-free spraying as a method for antisolvent application (**Figure 1b**). In this simple method, the antisolvent is introduced into a vessel with a thin tube connected to a pump. Upon pressing on the pump trigger, the antisolvent is forced through a one-way valve into a nozzle, creating a mist of antisolvent droplets that land on the wet perovskite film. This method allows to dispense the antisolvent very quickly and in fixed quantities, unlike the continuous spraying that occurs in the carrier-gas-assisted method. The overall amount of antisolvent interacting with the wet perovskite film can be tuned, by controlling the number of spray cycles and the height from which the antisolvent is sprayed. We focus on the fabrication of stoichiometric triple-cation mixed-halide perovskite ($Cs_{0.05}(MA_{0.17}FA_{0.83})_{0.95}Pb(I_{0.9}Br_{0.1})_3$) films and compare the deposition of different volumes of antisolvent (60 µl and 200 µl) by either pipetting or carrier-gas-free spraying. We find that solar cells fabricated when spraying the antisolvent outperform those in which the antisolvent was pipetted, reaching a maximum power conversion efficiency (PCE) of 21%. Scanning electron microscopy (SEM) and grazing incidence wide angle x-ray scattering (GIWAXS) studies revealed that the sprayed samples exhibit a strongly reduced $PbI_2$ content as compared to the pipetted ones, suggesting a much better preservation of the intended stoichiometry of the perovskite films. This leads to an improvement in the open-circuit voltage ($V_{OC}$) of the devices and an improved shelf-stability.

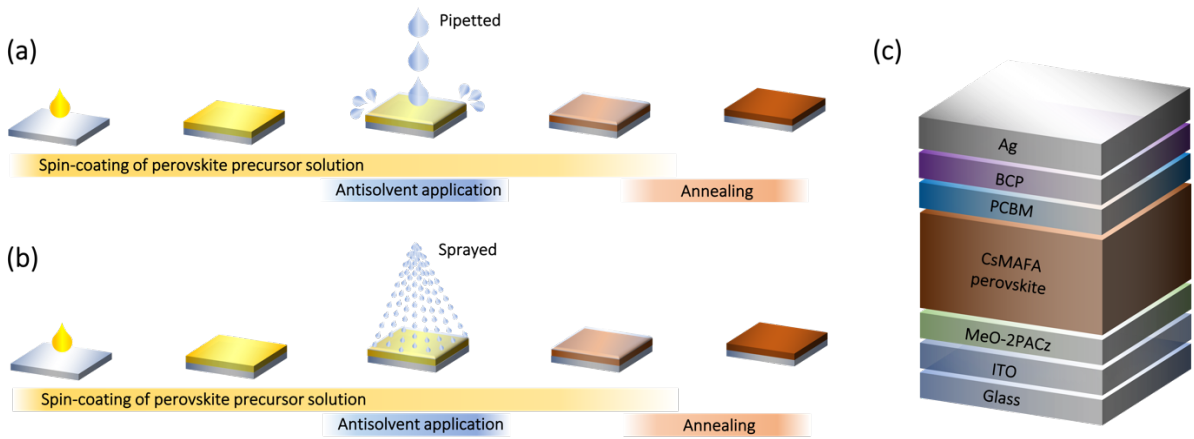

**Figure 1.** Schematic depiction of the perovskite layer fabrication process via (a) pipetted and (b) sprayed antisolvent application. (c) Schematic structure of the photovoltaic devices fabricated in this work.

**Photovoltaic device performance.**

To investigate the efficacy of carrier-gas-free antisolvent spraying in the formation of high-quality perovskite layers, we fabricated solar cells in an inverted architecture, the structure of which is shown in **Figure 1c**. The photovoltaic parameters of the best 24 solar cells of each kind are presented in **Figure 2**. For both volumes of antisolvent (60 or 200 μL of trifluorotoluene, TFT) the $V_{OC}$ of the sprayed devices is on average 20-30 mV higher than for the pipetted ones. The short-circuit current ($J_{SC}$) of the sprayed devices also shows a minor increase, especially evident for the low volume of antisolvent. The fill factors (FF) remain essentially unchanged, with most devices resulting in FF>80%. These improvements lead to a higher PCE of the sprayed devices, with an average PCE >20% and a maximum PCE of 21%. The current-voltage (J-V) characteristics of the best-performing devices of each kind are shown in **Figure S1** and their photovoltaic performance parameters are summarized in **Table S1**. While both pipetted and sprayed devices exhibit negligible hysteresis, it is noteworthy that the hysteresis of the sprayed devices is particularly low. Interestingly, increasing the antisolvent volume led to an increase in the performance of the pipetted devices – in agreement with earlier observations[15] – but the performance of the sprayed devices was similarly good for both 60 μL and 200 μL of antisolvent. We emphasize that the architecture of the devices, the extraction layers and all other fabrication parameters were kept the same, so the improvement in the photovoltaic performance is associated solely with the method of antisolvent deposition.

Since the performance improvement originates mainly from the changes in $V_{OC}$, we explored whether this variance arises from a different bandgap of the perovskite layers. UV-vis measurements (**Figure 3a**) confirmed that regardless of the antisolvent application method, the bandgap of the sprayed/pipetted perovskite layers remains unchanged. Another possibility for the increase in $V_{OC}$ is that the non-radiative losses are suppressed for devices with sprayed antisolvent.[26] To explore that, we measured the photoluminescence (PL) of complete photovoltaic devices of each kind (**Figure 3b**). We find that the PL of sprayed devices is significantly higher than that of the pipetted ones. Consequently, their PL quantum efficiency (PLQE) is more than doubled as compared to the pipetted solar cells (see inset). The enhanced PLQE is consistent with the observed increase in $V_{OC}$, suggesting that perovskite films fabricated with sprayed antisolvent exhibit a lower density of non-radiative recombination centers.

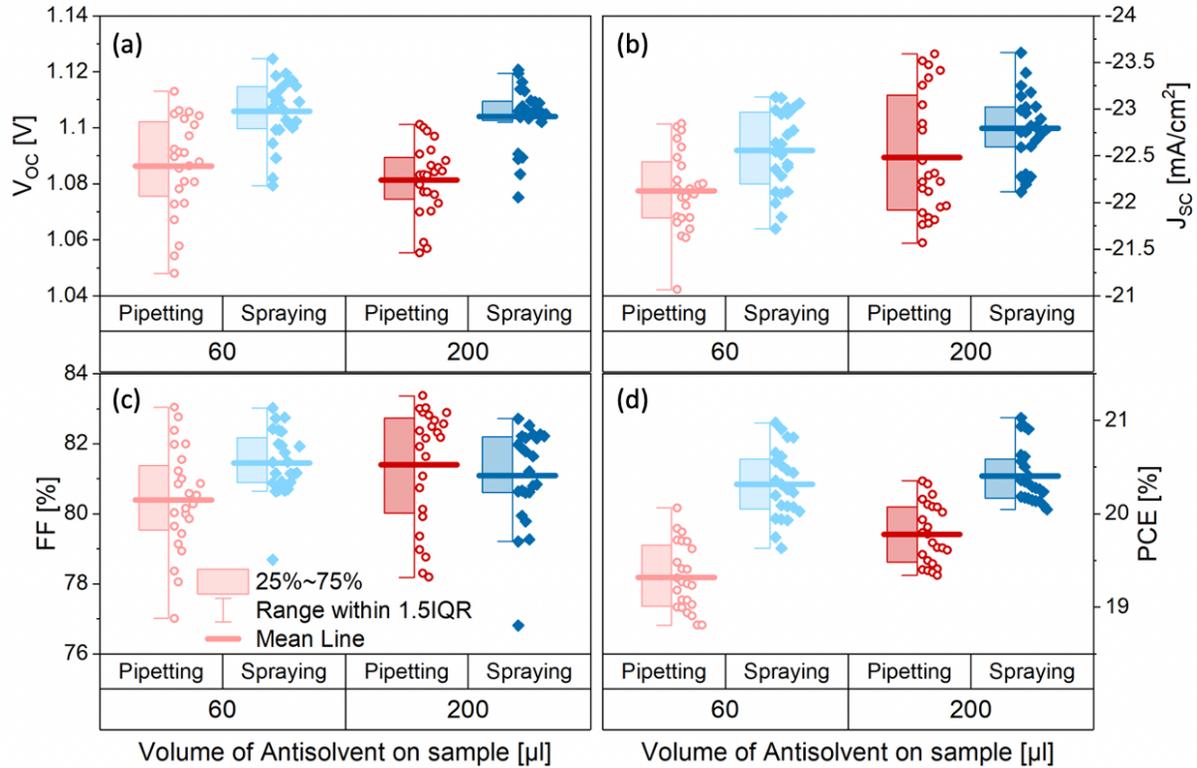

**Figure 2**. (a) $V_{OC}$ (b) $J_{SC}$ (c) FF and (d) PCE of triple cation perovskite solar cells fabricated by either spraying (blue diamonds) or pipetting (red circles) 60 or 200 μL of TFT as antisolvent.

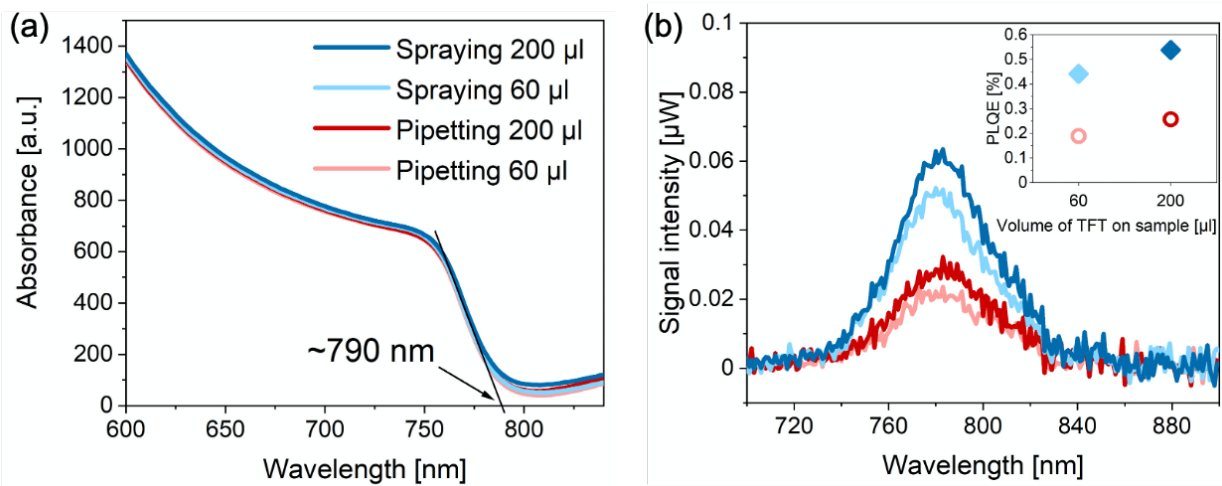

**Figure 3**. (a) Absorption spectra of triple cation perovskite thin films and (b) photoluminescence spectra and photoluminescence quantum efficiency (PLQE) of triple cation perovskite solar cells fabricated by either spraying or pipetting either 60/200 μL of TFT.

**Structural characterization of pipetted and sprayed perovskite films.**

To investigate the microstructure of perovskite films fabricated by either pipetting or spraying the antisolvent, the films were characterized by SEM. All films show the typical compact polycrystalline microstructure with grains ranging from 100 nm to 400 nm in size. In the case of films fabricated by pipetting the antisolvent (**Figure 4a** and **4b**), we also observe multiple crystallites at the surface of the perovskite film (marked in red) that are commonly associated with the presence of phase-separated $PbI_2$ phase in the perovskite film. On the other hand, samples fabricated by spraying the antisolvent (**Figure 4c** and **4d**) show significantly fewer of such $PbI_2$ crystallites.

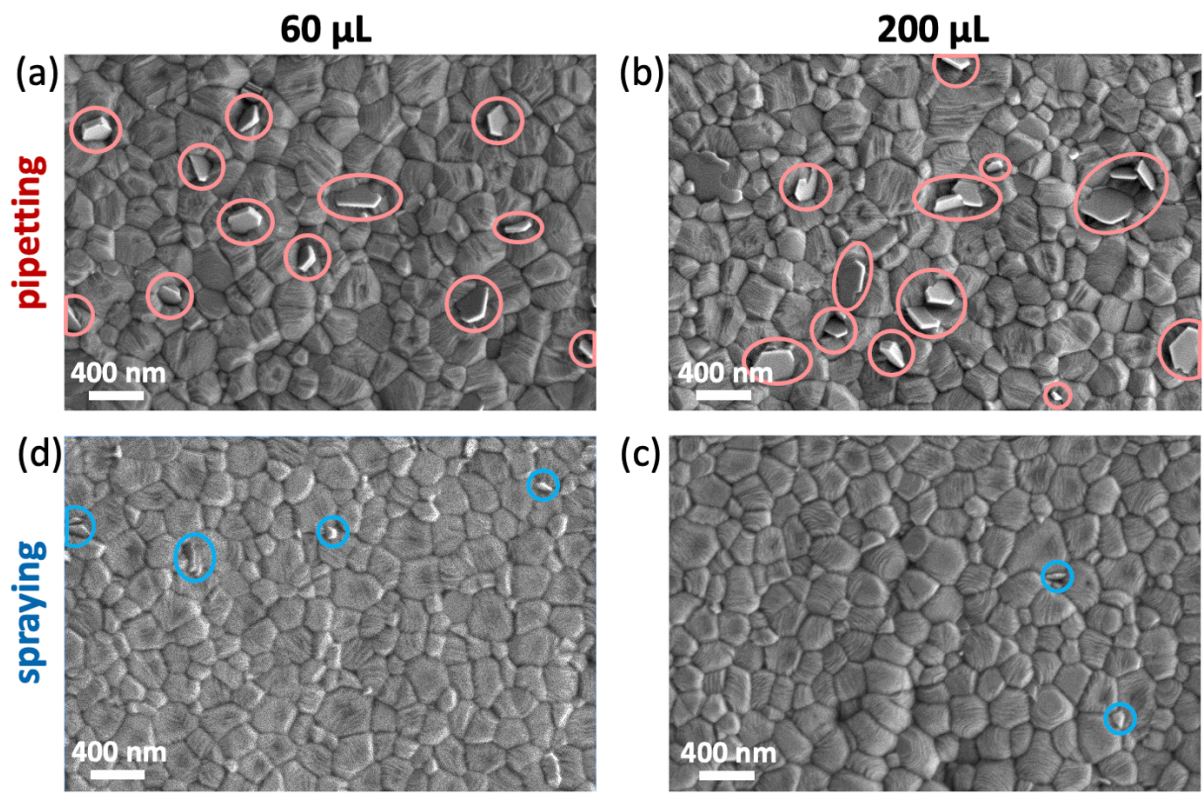

**Figure 4**. Scanning electron microscopy (SEM) images of triple cation perovskite thin films fabricated by either (a/b) pipetting or (c/d) spraying 60/200 μL of TFT.

To further investigate the crystalline structure of the perovskite films, they were characterized by GIWAXS (**Figure 5a-d**) at the surface scattering instrument of ID10 beamline of the European Synchrotron Radiation Facility (ESRF), Grenoble (France). The experimental configuration of the GIWAXS measurements was similar to that in earlier works.[27-29] In all cases, the samples exhibit the typical triple-cation crystal structure (Supplementary Information, **Figure S2**) with a small contribution of excess $PbI_2$ phase. Quantitatively, the $PbI_2$ volume fraction in all samples is larger at the surface compared to the bulk of the films. Furthermore, samples formed by pipetting the antisolvent exhibit a higher $PbI_2$ content than sprayed samples at both the surface and the bulk of the perovskite layers (**Figure 5e**).

Taken together with the SEM images, these results suggest that spraying the antisolvent far better preserves the intended stoichiometry of the perovskite layers. We note that unlike many reports, in which excess $PbI_2$ is intentionally introduced into the perovskite composition,[30-31] the intended composition of the perovskite films in our study is perfectly stoichiometric. Consequently, the presence of $PbI_2$ domains suggests that a certain fraction of organic halides was absent during the crystallization of the perovskite layer. Such a loss of organic halides is likely to occur during the application of an antisolvent, even for antisolvents that do not easily dissolve organic halides, such as TFT.

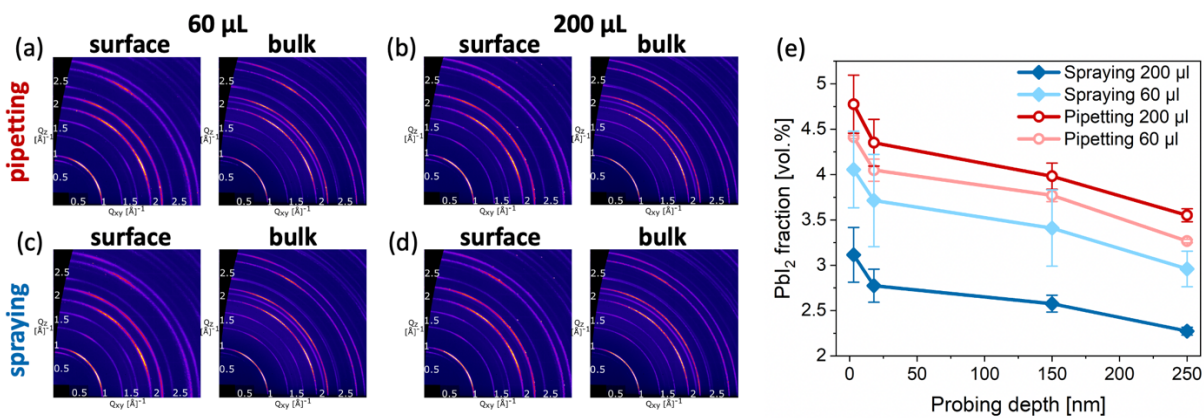

**Figure 5**. GIWAXS measurements of triple cation perovskite thin films fabricated by either (a/b) pipetting or (c/d) spraying 60/200 μL of TFT (e) volume fraction of $PbI_2$ at different depths of the different perovskite films.

Considering the differences between the two antisolvent application methods offers an explanation for the variation in the final composition of the perovskite layers. When dispensing the antisolvent by pipetting, its large volume comes in contact with the wet perovskite film, allowing the diffusion of organic halides such as methylammonium iodide (MAI) and formamidinium iodide (FAI) from the wet perovskite film into the antisolvent, which due to the high centrifugal forces is spun off the surface of the sample (**Figure 6**). With a certain fraction of organic halides lost, the remaining precursors crystallize into a perovskite film that contains a certain amount of PbI$_2$. The impact is strongest at the surface of the perovskite film, but is evident also in its bulk. On the other hand, by spraying the antisolvent, it interacts with the surface of the sample as a mist of small droplets. The relatively small volume of each droplet significantly lowers the maximum possible amount of organic halides that can diffuse into the antisolvent. Moreover, the large surface area of the sprayed droplets results in an accelerated evaporation, which is significantly more rapid than that of the pipetted antisolvent volume. Finally, the droplets are less likely to be spun off the surface of the sample due to their lower mass. All these factors contribute to the far better preservation of the intended perovskite composition in the final crystallized film.

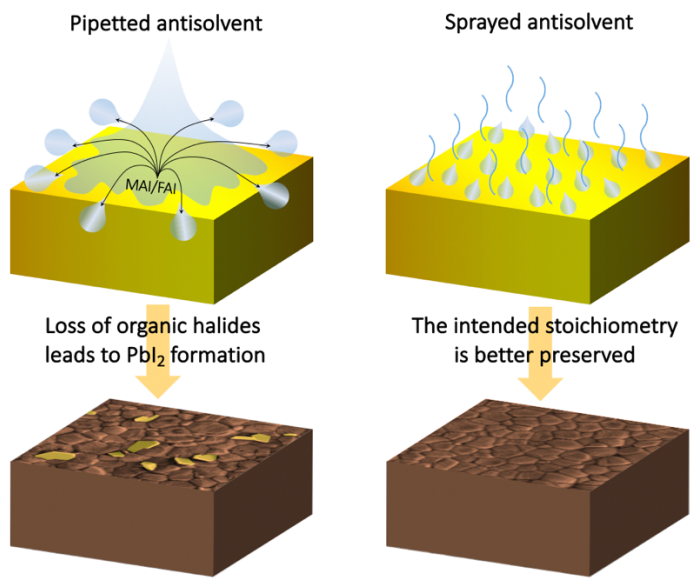

**Figure 6**. Schematic presentation of the differences between pipetting and spraying the antisolvent and the resultant film stoichiometry.

The presence of larger amounts of PbI$_2$ in the pipetted perovskite films results not only in a reduced initial photovoltaic performance, but also in a reduced device stability. We tracked the performance of unencapsulated solar cells fabricated with either spraying or pipetting the antisolvent over a period of two months (**Figure S3**). Between measurements, the devices were kept in darkness and ambient air. While both types of devices show a significant decrease in performance, we find that the sprayed devices are generally more stable than the pipetted ones. This is consistent with previous reports, showing that unreacted PbI$_2$ can have detrimental effects on the stability of perovskite solar cells,[32] and that its removal by various surface modification methods can lead to improved performance and stability.[33]

Our results indicate that the formation of unreacted PbI$_2$ can be largely avoided by application of the antisolvent by carrier-gas free spraying rather than pipetting. This enables the fabrication of high efficiency devices using a relatively low volume of antisolvent (60 μL). Considering the simplicity of the spraying method, which requires no additional setups or equipment, we believe this method can be widely adopted by other researchers. Moreover, these results also open the route for fabricating more efficient and stable devices in large-area production where spraying is particularly suitable.

**Supporting Information**. Experimental information, J-V characteristics and photovoltaic performance parameters, XRD 1D profiles, device stability data.

Experimental data from the X-ray scattering data: [DOI10.15151/ESRF-ES-406842143](DOI10.15151/ESRF-ES-406842143)

AUTHOR INFORMATION

Corresponding author:

Prof. Dr. Yana Vaynzof

Email address: yana.vaynzof@tu-dresden.de

URL of the group website: https://cfaed.tu-dresden.de/cfeet-about

Twitter handle: @vaynzof

ACKNOWLEDGMENT

This work is financially supported by the European Research Council (ERC) under the European Union's Horizon 2020 research and innovation program (ERC Grant Agreement No. 714067, ENERGYMAPS). We acknowledge the Deutsche Forschungsgemeinschaft (DFG) for funding the <PROCES> project (Grant No. VA 991/2-1) and the <PERFECT PVs> project (Grant No. 424216076) and the <PHIVE-X> project (SCHR 700/38-1) in the framework of SPP 2196. We also thank the BMBF for funding (project 05K19VTA). Finally, we kindly acknowledge the European Synchrotron Radiation Facility for provision of synchrotron radiation facilities and the Dresden Center for Nanoanalysis in the cfaed for providing access to the electron microscopy facilities.